\documentclass[aps]{revtex4}%
\usepackage{amsfonts}
\usepackage{amsmath}
\usepackage{amssymb}
\usepackage{graphicx}%
\begin{document}
\title[Bound electron states in impure graphene]{Bound electron states in the monolayer graphene with the short-range impurities}
\author{Natalie E. Firsova}
\affiliation{Institute for Problems of Mechanical Engineering, the
Russian Academy of Sciences, St. Petersburg 199178, Russia}
\author{Sergey A. Ktitorov}
\affiliation{A.F. Ioffe Physical-Technical Institute, the Russian
Academy of Sciences, St. Petersburg, Russia}
\author{Philip A. Pogorelov}
\affiliation{St. Petersburg State University, Pervogo Maya str.
100, Petrodvoretz, St. Petersburg 198504, Russia} 

\begin{abstract}
Bound electron states in impure graphene are considered. Short-range
perturbations for defect and impurities of the types "local chemical
potential" and "local gap" are taken into account.

\end{abstract}
\maketitle

%\preprint{HEP/123-qed}

%\pacs{PACS number}

%\volumeyear{year}
%\volumenumber{number}
%\issuenumber{number}
%\eid{identifier}
%\date[Date text]{date}
%\received[Received text]{date}

%\revised[Revised text]{date}

%\accepted[Accepted text]{date}

%\published[Published text]{date}

%\startpage{101}
%\endpage{102}
%\tableofcontents

\section{Introduction}

The Dirac equation is a fundamental base of the relativistic field
theory. However, it is an important model in the non-relativistic
solid state theory as well. Superconductors with $d-$pairing
\cite{d}, the Cohen-Blount two-band model of narrow-gap
semiconductors \cite{keldysh}, \cite{tamar}, electronic spectrum
of the carbon tubes form an incomplete list of the
non-relativistic applications of this equation. During the last
two years extremely much attention was payed to the problem of the
electronic spectrum of graphene (see for the review
\cite{novosel}). Two-dimensional structure of it and a presence of
the cone points in the electronic spectrum make actual a
comprehensive study of the external fields effect on the spectrum
and other characteristics of the electronic states described by
the Dirac equation in the 2+1 space-time. We consider in this work
the bound states of the 2+1 Dirac equation due to the short-range
perturbation. Particular attention to this case stems from the
effectiveness of short-range scatterers in contrast to the
long-range ones: an effect of the latter is suppressed by the
Klein paradox \cite{beenakker}. Our work takes into account the
obvious fact that the Kohn-Luttinger matrix elements of the
short-range perturbation calculated on the upper and lower band
wave functions are not equal in a general case. This means that in
the perturbed Dirac equation not only the potential but the mass
perturbation can be present.

\section{Perturbed Dirac equation in (2+1)-space-time}

The Dirac equation describing electronic states in graphene reads
\cite{novosel}
\begin{equation}
\left(  -is\hbar\sum_{\mu=1}^{2}\sigma_{\mu}\partial_{\mu}-\sigma_{3}\left(
m+\delta m\right)  s^{2}\right)  \psi=\left(  E-V\right)  \psi,
\label{diracgeneral}%
\end{equation}
where $s$ is the limiting velocity of the band electrons,
$\sigma_{\mu}$ are the Pauli matrices, $2ms^{2}=E_{g}$ is the
electronic spectrum gap, $\psi\left(  \mathbf{r}\right)  $ is the
two-component spinor. The spinor structure takes into account the
two-band nature.$\ \delta m\left( \mathbf{r}\right)  $ and
$V(\mathbf{r})$ are the local perturbations of the mass (gap) and
the chemical potential. A local mass perturbation can be induced
by defects in the graphene film or in the substrate \cite{gap}. We
consider here the delta function model of the perturbation:%
\begin{equation}
\delta m\left(  \mathbf{r}\right)  =-b\delta(r-r_{0}),\text{
}V(\mathbf{r)}%
=-a\delta(r-r_{0}), \label{delta}%
\end{equation}
where $r$ and $r_{0}$ are respectively the polar coordinate radius and the
perturbation radius. Such short-range perturbation (and the equivalent form
$diag(V_{1},V_{2})\delta(r-r_{0})$ with $-V_{1}=\frac{a+b}{2},-V_{2}%
=\frac{a-b}{2}$) was used in the (3+1)-Dirac problem for
narrow-gap and zero-gap semiconductors in \cite{tamar}. The
two-dimensional Dirac problem with the scalar short-range
perturbation (\ref{delta}) (but without the mass perturbation) was
considered in \cite{dong}. The obtained there characteristic
equation for the discrete spectrum energy contains one mistake. We
correct it here and take account of the mass perturbation $\delta
m\left(  \mathbf{r}\right)  .$

Let us present the two-component spinor in the form%
\begin{equation}
\psi_{j}(\mathbf{r},t)=\frac{\exp\left(  -iEt\right)  }{\sqrt{r}}%
\begin{pmatrix}
f_{j}\left(  r\right)  \exp\left[  i\left(  j-1/2\right)  \phi\right] \\
g_{j}\left(  r\right)  \exp\left[  i\left(  j+1/2\right)  \phi\right]
\end{pmatrix}
, \label{spinor}%
\end{equation}
where $j$ is the pseudospin quantum number; $j=\pm1/2,\pm3/2,\ldots$. In the
opposite to the relativistic theory, this quantum number has nothing to do
with the real spin and indicates the degeneracy in the biconic Dirac point.
The upper $f_{j}\left(  r\right)  $ and $g_{j}\left(  r\right)  $ components
of the spinor satisfy the equations
\begin{equation}
\frac{dg_{j}}{dr}+\frac{j}{r}g_{j}-\left(  E-m\right)  f_{j}=\left(
a+b\right)  \delta(r-r_{0})f_{j}, \label{componenteq1}%
\end{equation}

\begin{equation}
-\frac{df_{j}}{dr}+\frac{j}{r}f_{j}-\left(  E+m\right)  g_{j}=\left(
a-b\right)  \delta(r-r_{0})g_{j}. \label{componeq2}%
\end{equation}
These equations have a symmetry:
\begin{equation}
f_{j}\leftrightarrow g_{j},\text{ }E\rightarrow-E,\text{
}j\rightarrow-j \label{symm}%
\end{equation}
Let us introduce the function $\varphi_{j}\left(  r\right)  \equiv
f_{j}/g_{i}.$ It satisfies the equation:%
\begin{equation}
\frac{1}{\left(  a+b\right)  \varphi_{j}^{2}+\left(  a-b\right)  }\left[
\frac{d\varphi_{j}}{dr}-\frac{2j}{r}\varphi_{j}-E\left(  \varphi_{j}%
^{2}+1\right)  \right]  +\delta(r-r_{0})=0 \label{phi}%
\end{equation}
Integrating in the vicinity of $r=r_{0}$%
\begin{equation}
\lim_{\epsilon\rightarrow0}\int_{\varphi_{j}(r_{0}-\epsilon)}^{\varphi
_{j}(r_{0}+\epsilon)}\frac{d\varphi_{j}}{\left(  a+b\right)  \varphi_{j}%
^{2}+\left(  a-b\right)  }=-1, \label{match1}%
\end{equation}
we obtain the matching condition%
\begin{equation}
\arctan\left(  \varphi_{j}^{-}\sqrt{\frac{a+b}{a-b}}\right)  -\arctan\left(
\varphi_{j}^{+}\sqrt{\frac{a+b}{a-b}}\right)  =\sqrt{a^{2}-b^{2}},
\label{match2}%
\end{equation}
where $\varphi_{j}^{-}\equiv\varphi_{j}\left(  r_{0}-\epsilon\right)
,\varphi_{j}^{+}\equiv\varphi_{j}\left(  r_{0}+\epsilon\right)  ,a^{2}>b^{2}.$
The upper and lower component matching condition resulting from (\ref{match2})
reads%
\begin{equation}
\left(
\begin{array}
[c]{c}%
f_{j}^{+}\\
g_{j}^{+}%
\end{array}
\right)  =\overset{\wedge}{A}\left(
\begin{array}
[c]{c}%
f_{j}^{-}\\
g_{j}^{-}%
\end{array}
\right)  , \label{matrixrelation}%
\end{equation}

where the matrix $\overset{\wedge}{A}$
\begin{equation}
\left(
\begin{array}
[c]{cc}%
\cos\sqrt{a^{2}-b^{2}}, & -\sqrt{\frac{a-b}{a+b}}\sin\sqrt{a^{2}-b^{2}}\\
\sqrt{\frac{a-b}{a+b}}\sin\sqrt{a^{2}-b^{2}}, & \cos\sqrt{a^{2}-b^{2}}%
\end{array}
\right)  \label{matrix}%
\end{equation}

is orthogonal for $b=0.$ It transmutes into the matrix%
\begin{equation}
\left(
\begin{array}
[c]{cc}%
\cosh\sqrt{b^{2}-a^{2}}, & -\sqrt{\frac{b-a}{b+a}}\sinh\sqrt{b^{2}-a^{2}}\\
\sqrt{\frac{b-a}{b+a}}\sinh\sqrt{b^{2}-a^{2}}, & \cosh\sqrt{b^{2}-a^{2}}%
\end{array}
\right)  , \label{matrix2}%
\end{equation}

when $a^{2}-b^{2}<0.$

The general solution can be found solving the second-order
equation obtained by excluding one of the spinor components from
the equation set (\ref{componenteq1}), (\ref{componeq2}) in the
domains $0<r<r_{0}$ and
$r>r_{0}:$%
\begin{equation}
\frac{d^{2}f_{j}}{dr^{2}}+\left[   E^{2}-m^{2}-\frac{j\left(
j-1\right)  }{r^{2}}  \right]  f_{j}=0. \label{secondorder}%
\end{equation}
This equation is related to the Bessel one. Its general solution reads%
\begin{equation}
f_{j}=C_{1}\sqrt{r}I_{j-1/2}\left(  \kappa r\right)  +C_{2}\sqrt{r}%
K_{j-1/2}\left(  \kappa r\right)  , \label{general}%
\end{equation}
where $\kappa^{2}=m^{2}-E^{2},$ $I_{\nu}\left(  z\right)  $ and $K_{\nu
}\left(  z\right)  $ are the modified Bessel functions. The constant $C_{2}=0$
in the domain $0<r<r_{0}$, while $C_{1}=0$ in the domain $r>r_{0.}$ Expressing
the $g_{j}$-component using (\ref{componeq2}), we can write%
\begin{equation}
\varphi_{j}^{-}=\sqrt{\frac{m+E}{m-E}}\frac{I_{j-1/2}\left(  \kappa
r_{0}\right)  }{I_{j+1/2}\left(  \kappa r_{0}\right)  }, \label{phi+}%
\end{equation}

\begin{equation}
\varphi_{j}^{+}=\sqrt{\frac{m+E}{m-E}}\frac{K_{j-1/2}\left(  \kappa r\right)
}{K_{j+1/2}\left(  \kappa r\right)  }. \label{phi-}%
\end{equation}

Applying the matching condition (\ref{match2}) to the expressions
(\ref{phi-}), (\ref{phi+}) we obtain the characteristic equation for the bound
state energy levels:%

\[
\kappa\left[  \frac{K_{j-1/2}\left(  \kappa r_{0}\right)  }{K_{j+1/2}\left(
\kappa r_{0}\right)  }-\frac{I_{j-1/2}\left(  \kappa r_{0}\right)  }%
{I_{j+1/2}\left(  \kappa r_{0}\right)  }\right]  =
\]

\begin{equation}
-\frac{\tan\left(  \sqrt{a^{2}-b^{2}}\right)  }{\sqrt{a^{2}-b^{2}}}\left[
(m-E)\left(  a-b\right)  +\left(  a+b\right)  (m+E)\frac{I_{j-1/2}\left(
\kappa r_{0}\right)  }{I_{j+1/2}\left(  \kappa r_{0}\right)  }\frac
{K_{j-1/2}\left(  \kappa r_{0}\right)  }{K_{j+1/2}\left(  \kappa r_{0}\right)
}\right]  \label{character1}%
\end{equation}

where $a^{2}-b^{2}>0.$ This equation turns to the characteristic equation
obtained in \cite{dong}, for $b=0$ apart from the mistakenly omitted terms in
the right hand side of (\ref{character1}). In the opposite case of
$a^{2}-b^{2}<0$ we have
\[
\kappa\left[  \frac{K_{j-1/2}\left(  \kappa r_{0}\right)  }{K_{j+1/2}\left(
\kappa r_{0}\right)  }-\frac{I_{j-1/2}\left(  \kappa r_{0}\right)  }%
{I_{j+1/2}\left(  \kappa r_{0}\right)  }\right]
=
\]%
\begin{equation}
-\frac{\tanh\left( \sqrt{b^{2}-a^{2}}\right)
}{\sqrt{b^{2}-a^{2}}}\left[  -(m-E)\left(  b-a\right)  +\left(
b+a\right)  (m+E)\frac {I_{j-1/2}\left(  \kappa r_{0}\right)
}{I_{j+1/2}\left(  \kappa r_{0}\right) }\frac{K_{j-1/2}\left(
\kappa r_{0}\right)  }{K_{j+1/2}\left(  \kappa
r_{0}\right)  }\right]  \label{character2}%
\end{equation}

We write these equations in another form making the symmetry (\ref{symm}) manifest:%

\[
\kappa\left[  I_{j-1/2}\left(  \kappa r_{0}\right)  K_{j+1/2}\left(  \kappa
r_{0}\right)  -K_{j-1/2}\left(  \kappa r_{0}\right)  I_{j+1/2}\left(  \kappa
r_{0}\right)  \right]  =
\]

\begin{equation}
\frac{\tan\left(  \sqrt{a^{2}-b^{2}}\right)  }{\sqrt{a^{2}-b^{2}}}\left[
(m-E)\left(  a-b\right)  I_{j+1/2}\left(  \kappa r_{0}\right)  K_{j+1/2}%
\left(  \kappa r_{0}\right)  +\left(  a+b\right)  (m+E)I_{j-1/2}\left(  \kappa
r_{0}\right)  K_{j-1/2}\left(  \kappa r_{0}\right)  \right]  ,
\label{character1a}%
\end{equation}

\[
\kappa\left[  I_{j-1/2}\left(  \kappa r_{0}\right) K_{j+1/2}\left(
\kappa r_{0}\right)  -K_{j-1/2}\left(  \kappa r_{0}\right)
I_{j+1/2}\left(  \kappa r_{0}\right)  \right] =
\]%
\begin{equation}
\frac{\tanh\left(  \sqrt{b^{2}-a^{2}}\right)  }%
{\sqrt{b^{2}-a^{2}}}\left[  -(m-E)\left(  b-a\right)
I_{j+1/2}\left(  \kappa r_{0}\right)
K_{j+1/2}\left(  \kappa r_{0}\right)  +\left(  b+a\right)  (m+E)I_{j-1/2}%
\left(  \kappa r_{0}\right)  K_{j-1/2}\left(  \kappa r_{0}\right)  \right]
\label{character2a}%
\end{equation}

\section{Analysis of the characteristic equation and numerical results}

Making use of the Bessel functions limiting forms for small arguments
\cite{stegun}
\[
I_{\nu}(z)\sim\left(  z/2\right)  ^{\nu}\frac{1}{\Gamma\left(  \nu+1\right)
},\text{ }K_{0}\left(  z\right)  \sim-\ln z,\text{ }K_{\nu}\left(  z\right)
\sim\frac{1}{2}\Gamma\left(  \nu\right)  \left(  z/2\right)  ^{-\nu},
\]
we can obtain a simple relation describing the asymptotic behaviour of the
energy level, where the perturbation power approaches zero:%
\begin{equation}
E=m\left[  1-\frac{r_{c}^{2}}{2r_{0}^{2}}\exp\left(  -\frac{r_{c}}%
{r_{0}\left(  a+b\right)  }\right)  \right]  , \label{approach}%
\end{equation}
where $r_{c}=m^{-1}$ (in units with $\hbar=s=1),$ $a+b>0.$ This
result conforms the well known general property of the
two-dimensional quantum systems: a threshold for creation of the
bound state is absent; the point $a+b=0$ is the essentially
singular point of the function $E=E(a+b).$ One can see that the
function $E(a)$ approaches the point $E=-m$ at some large enough
value of $a>0$. $.$ Making use of the Bessel function asymptotic
behaviour \cite{stegun},
\[
I_{\nu}\left(  z\right)  \sim\left(  2\pi z\right)  ^{-1/2}\exp z,\text{
}K_{\nu}\left(  z\right)  \sim\left(  \frac{\pi}{2z}\right)  ^{1/2}\exp\left(
-z\right)  ,
\]
and the equation \ref{character2}, \ we can see that the function $E(b)$
approaches the point $E=0$ when $\frac{r_{0}}{r_{c}}$ is large enough and
$b\rightarrow\infty$.

In the Fig. 1 the electron bound state energy is presented as a function of
the potential amplitude for the angular momentum quantum number $j=1/2$,
$\frac{r_{0}}{r_{c}}=1$ and $b=0.$ Inspecting this plot one can see that our
analytic solution (\ref{approach}) perfectly approximates approaching of the
bound state energy value the upper band bottom, when $b$ approaches zero.

In the Fig. 2 the bound state energy is presented as a function of the mass
perturbation amplitude $b$ for $a=0,$ $\frac{r_{0}}{r_{c}}=1,$ $j=1/2.$

In the Fig. 3 the electron bound state energy is presented as a
function of the potential amplitude for the angular momentum
quantum number $j=1/2$, $\frac{r_{0}}{r_{c}}=1,$ and $b=-1.$ We
see that the energy dependence on $a$ is non-monotonic function,
but approaching the upper band bottom takes place similarly to the
case of $b=0.$

\section{Conclusion}

In conclusion, we considered the bound electron states for the
two-dimensional Dirac equation with the short-range perturbation.
The short-range perturbation is approximated by the delta function
$\delta\left(  r-r_{0}\right)  $ with different amplitudes in the
upper and lower bands. We found the characteristic equation for
the discrete energy levels. Energy levels behaviour in dependence
on the perturbation amplitudes was investigated both analytically
and numerically. The obtained results can be useful for
understanding of the graphene electron properties.

%BeginExpansion
\begin{figure}
[tb]
\begin{center}
\includegraphics[
height=3.0364in, width=4.088in
]%
{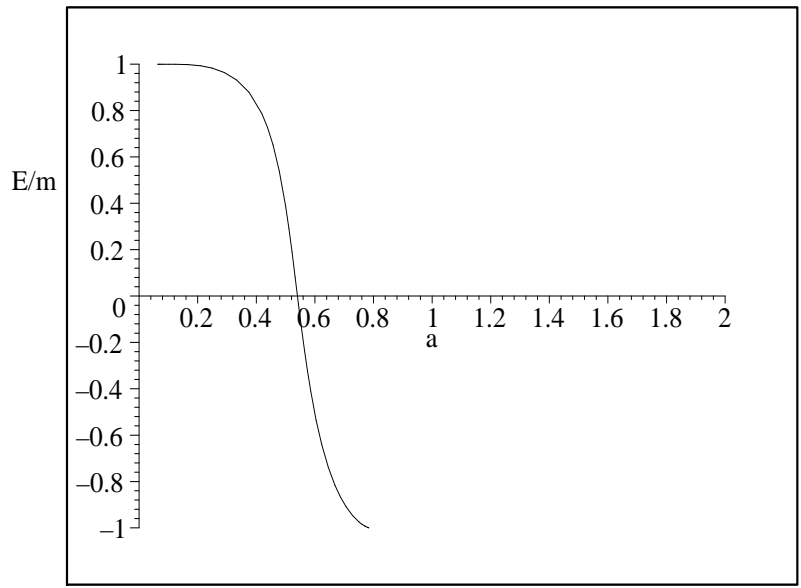}%
\caption{Reduced lower electron bound state energy E/m dependence
on the
short-range potential amplitude a at b=0. }%
\end{center}
\end{figure}
%EndExpansion

%BeginExpansion
\begin{figure}
[t]
\begin{center}
\includegraphics[
height=3.0364in, width=4.088in
]%
{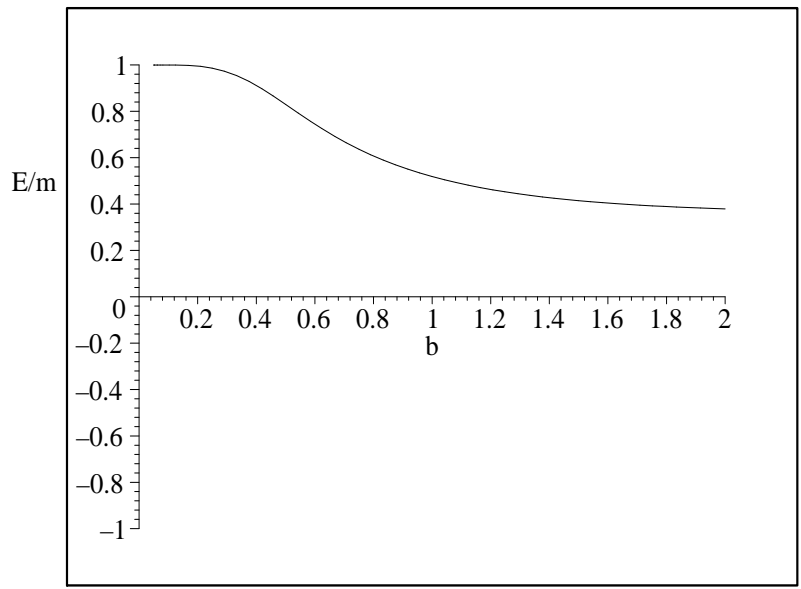}%
\caption{Reduced lower electron bound state energy E/m dependence
on the mass
perturbation amplitude b at a=0. }%
\end{center}
\end{figure}
%EndExpansion

%BeginExpansion
\begin{figure}
[t]
\begin{center}
\includegraphics[
height=3.0364in, width=4.088in
]%
{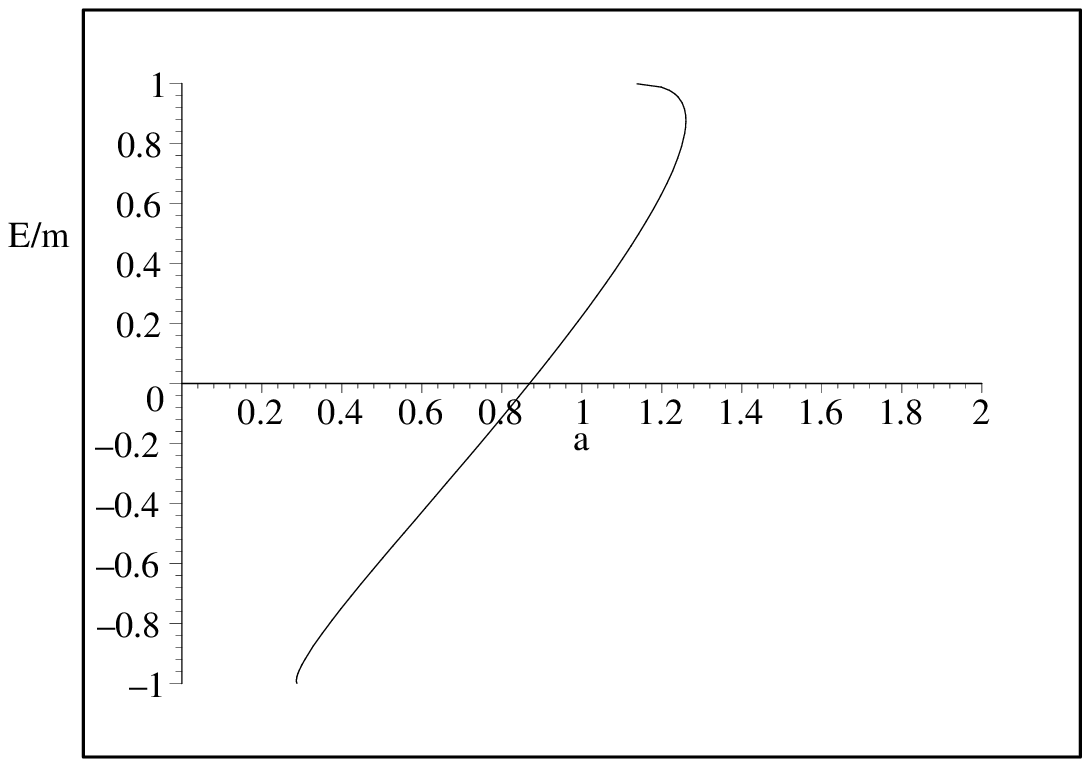}%
\caption{Reduced lower electron bound state energy E/m dependence
on the
short-range potential amplitude a at b=-1. }%
\end{center}
\end{figure}
%EndExpansion


\begin{thebibliography}{9}                                                                                                %


\bibitem {d} D.J. Scalapino, Phys. Rep. \textbf{250}, 329 (1995).

\bibitem {keldysh} L.V. Keldysh, JETP \textbf{45}, 365 (1963).

\bibitem {tamar}S.A. Ktitorov, V.I. Tamarchenko, Soviet Physics (Solid State)
\textbf{19}, 2070 (1977).

\bibitem {novosel}A.H. Castro Neto, F. Guinea, et al, arXiv: 0709.1163 (2008)
(accepted to Rev. Mod. Phys).

\bibitem {beenakker}C.W.J. Beenakker, Rev. Mod. Phys., \textbf{80}, 1337 (2008).

\bibitem {gap}Aurelien Lherbier, X. Blaze, et al, Phys. Rev. Letters,
\textbf{101}, 036808-1 (2008).

\bibitem {dong}Shi-Hai Dong, Zhong-Qi Ma, Phys. Lett. A \textbf{15}, 171 (2002).

\bibitem {stegun}M. Abramowitz, I.A. Stegun, \textit{Handboock of Mathematical
Functions with Formulas, Graphs, and Mathematical Tables, }National Bureau of
Standards, Washington DC, 1964.
\end{thebibliography}
\end{document}